\documentclass[fleqn,usenatbib,letters]{mnras}
\usepackage[british]{babel}

\usepackage{amssymb,multirow}
\usepackage{amsmath}
\usepackage{bm}
\usepackage{graphicx}
\usepackage{epstopdf}
\usepackage{caption}
\usepackage{float}
\usepackage{hyperref}

\newcommand{\nbs}{\nobreak\hspace{.16667em plus .08333em}}

\providecommand*{\aap}{{\it A\&A}}
\providecommand*{\mnras}{{\it MNRAS}}

\providecommand*{\grl}{{\it Geophys. Res. Lett.}}

\providecommand*{\pre}   {{\it Physical Review E}}
\providecommand*{\apj}   {{\it Astrophys. J.}}

\renewcommand{\vec}[1]{\bm{#1}}

\newcommand{\dif}[1]{\vec\nabla^2#1}

\newcommand{\ey}{\vec e_y}

\newcommand{\mr}[1]{\mathrm{#1}}

\def\rrr#1\\{\par
\medskip\hbox{\vbox{\parindent=2em\hsize=6.12in
\hangindent=4em\hangafter=1#1}}}

\usepackage{color}

\title[Dynamo in Shearing Box]{Magnetorotational Dynamo Action in the Shearing Box}

\author[J. Walker \& S. Boldyrev]{
Justin Walker$^{1}$\thanks{E-mail: jwwalker2@wisc.edu}
and Stanislav Boldyrev$^{1,2}$
\\
$^{1}$Department of Physics, University of Wisconsin-Madison, 1150 University Avenue, Madison, WI 53706, USA\\
$^{2}$Space Science Institute, Boulder, Colorado 80301, USA\\
}

\date{\today}

\pubyear{2017}

\begin{document}
\label{firstpage}
\maketitle

\begin{abstract}
  Magnetic dynamo action caused by the magnetorotational instability is studied in the shearing-box approximation with no imposed net magnetic flux.
  Consistent with recent studies, the dynamo action is found to be sensitive to the aspect ratio of the box: it is much easier to obtain in tall boxes (stretched in the direction normal to the disk plane) than in long boxes (stretched in the radial direction).
  Our direct numerical simulations indicate that the dynamo is possible in both cases, given a large enough magnetic Reynolds number.
  To explain the relatively larger effort required to obtain the dynamo action in a long box, we propose that the turbulent eddies caused by the instability most efficiently fold and mix the magnetic field lines in the radial direction.
  As a result, in the long box the scale of the generated  strong azimuthal (stream-wise directed) magnetic field is always comparable to the scale of the turbulent eddies.
  In contrast, in the tall box the azimuthal magnetic flux spreads in the vertical direction over a distance exceeding the scale of the turbulent eddies.
  As a result, different vertical sections of the tall box are permeated by large-scale nonzero azimuthal magnetic fluxes, facilitating the instability.
  In agreement with this picture, the cases when the dynamo is efficient are characterized by a strong intermittency of the local azimuthal magnetic fluxes.
\end{abstract}


\section{Introduction}
Magnetorotational instability (MRI), the instability of a differentially rotating, conducting fluid, permeated by a weak magnetic field \cite[][]{velikhov1959,chandra1960,balbus1991}, is believed to be the leading mechanism that renders astrophysical accretion flows turbulent and allows them to efficiently lose their angular momentum \cite[][]{balbus1998}.
The local properties of the instability and the resulting turbulence may be studied in the shearing-box approximation, which models a small box in the midplane of a disc so that the large-scale effects such as  stratification and stream-line curvature can be neglected \cite[][]{goldreich1965,hawley1995,umurhan2004,fromang2007,longaretti2010,lesur2011,riols_etal2015}.
The dynamics studied in the shearing box are therefore expected to be universal, that is, applicable to other astrophysical and laboratory systems possessing rotating, shearing flows \cite[e.g.,][]{kagan2014,petitdemange2008,ji2013}.

The properties of magnetic turbulence driven by the MRI in the shearing box are highly nontrivial.
In order to discuss them we use the following standard orthogonal coordinate frame.
Assuming that the box is positioned inside a disc, we chose the vertical coordinate ($z$) along the direction of the angular velocity, the coordinate ($x$) along the radial direction, and the azimuthal coordinate ($y$) along the unperturbed rotating flow.
When the box is permeated by a nonzero magnetic flux in the vertical direction, the linear instability develops, eventually leading to a strongly nonlinear steady-state turbulent regime.
Although the resulting turbulence is still not fully understood, it has been recently established that it resembles the standard magnetohydrodynamic turbulence \cite[][]{walker2016}.
The azimuthal magnetic field is concentrated at large scales and it plays the role of the guide field for small-scale fluctuations, whose energy spectrum and other statistical characteristic are close to that of homogeneous Alfv\'enic MHD  turbulence \cite[][]{zhdankin2017}.
A similar result has subsequently been obtained using hybrid (kinetic ions) plasma simulations \cite[][]{kunz2016}, which also found that sub-proton-scale MRI turbulence is consistent with the kinetic-Alfv\'en turbulence.
When the net vertical magnetic flux is zero, strong transient amplification is still possible if the net azimuthal magnetic flux is nonzero \cite[e.g.,][]{balbus1992,vishniac1992}.

In this work we concentrate on the case when the net magnetic flux through any cross-section of the box is zero.
In this zero-net-flux case, a linear instability is not possible, but the flow may become unstable for a finite initial perturbation (subcritical instability) \cite[][]{rincon2007,lesur2008,herault2011,riols_etal2013,squire2014,riols2016}.
Our recent high-resolution numerical simulations \cite[][]{walker2016} concentrated on the case of unit magnetic Prandtl number  ($\mr{Pm}$) and used the ``long" shearing box ($L_x:L_y:L_z=2:4:1$).
They did not find dynamo action for the magnetic Reynolds numbers several times larger than those required for the dynamo action in a non-shearing, non-rotating box.
This observation cast doubt on the existence of MRI dynamo action for $\mr{Pm}\leq 1$, in a stark contrast with the non-shearing, non-rotating case where the dynamo action is expected to exist for any given $\mr{Pm}$ as long as the magnetic Reynolds number is large enough \cite[e.g.,][]{boldyrev2004,boldyrev_etal2005,iskakov2007,boldyrev2010}.
A different outcome of the shearing-box dynamo simulations was however reported in \cite[][]{shi2016,nauman2016} who used the ``tall" shearing boxes, that is, boxes that vertical sizes exceed the radial sizes.
Quite interestingly, they observed that the critical Prandtl number for dynamo action\footnote{We use the term ``dynamo action'' to refer to the self-sustained turbulence that results from the subcritical MRI instability that, in turn, reinforces the magnetic field.} is reduced as long as $L_z/L_x\gtrsim 2.5$.

In order to understand the different outcomes of these studies, we have performed a series of direct numerical simulations of incompressible MHD dynamo action for varying aspect ratios of the shearing box.
We found that, in agreement with \cite[][]{shi2016} and \cite[][]{nauman2016}, the dynamo action is sensitive to the aspect ratio of the box.
In our case of $\mr{Pm}=1$, tall boxes (stretched in the $z$-direction) rapidly exhibit dynamo action.
To explain these results we propose that the turbulent eddies caused by the dynamo, efficiently fold the magnetic field lines in the radial ($x$) direction.
As a consequence, in  the long box the $x$-scale of the generated  $B_y$ component of the magnetic field is always comparable to the scale of the turbulent eddies.
In contrast, in the tall box the flux of $B_y$ can spread in the vertical direction over the distances exceeding the scales of the turbulent eddies, which are constrained by the short $x$~dimension of the box.
The vertical mixing of the $B_y$ field is thus suppressed in tall boxes.
As a result, different vertical sections of the tall box are permeated by large-scale nonzero fluxes of the azimuthal field $B_y$, leading to the instability.

This phenomenological picture motivated us to review one of the results from our previous work~\cite[][case IV]{walker2016}.
In this work, the dynamo action was not observed in the long box ($L_x:L_y:L_z=2:4:1$) for the long time during which we integrated the solution.
The initial large-scale fluctuations kept decaying for more than 200 shearing times.
During this decay, the scale of the turbulence kept decreasing as well, as to maintain the balance between the linear and turbulent shear rates.
For the large Reynolds number that we used, the turbulence would decay to progressively smaller scales.
In this case, however, the scale of the turbulent fluctuations should eventually become sufficiently smaller than the vertical extent of the box $L_z$, so that according to our phenomenological picture, the dynamo should become possible.
To check this hypothesis, we significantly extended the running time of simulations of \cite[][case IV]{walker2016}, and after about 600 shearing times did observe the possible onset of the dynamo action. 

This may reconcile the available numerical results.
Based on our findings, we suggest that the dynamo action is always possible, no matter what the aspect ratio of the box is.
However, the long boxes require significantly larger Reynolds number and (assuming large scale of the initial fluctuations) significantly longer running time in order to observe the dynamo action.
We have also established that the dynamo action leads to a quite intermittent distribution of the azimuthal magnetic fluxes, in agreement with the proposed phenomenological picture.    

\section{Numerical setup}
The shearing-box approximation captures the local small-scale dynamics of an accretion disc~\citep{goldreich1965}.
A small box centered at a radial distance $r_0$ and orbiting with the angular velocity $\Omega(r)$ has a shearing rate $q \equiv -( d \ln \Omega / d\ln r)_{r_0}$.
For the Keplerian orbital flow, we have $q=3/2$, and we denote $\Omega_0=\Omega(r_0)$.
At the center of the box $r_0$ the gravitational force balances the centrifugal force.
Assuming an incompressible flow, one writes the fluid momentum equation and the induction equation for the magnetic field in the vicinity of $r_0$, a detailed discussion of the derivation is given in~\cite[e.g.,][]{umurhan2004}.
The local orthogonal coordinate frame in the box is chosen such that the direction of $z$ coincides with the direction of $\Omega_0$, $x$ with the radial direction, and $y$ with the orbiting direction of the flow.
The equilibrium background flow in this box then corresponds to the constant linear shear $\vec{v}_0(x)=-q\Omega_0 x{\hat y}$.
By representing the velocity field as $\vec u = \vec v + \vec v_0$, we formally eliminate the background shear.
The shearing-box equations for the velocity and magnetic fields then take the form:\footnote{Because the mean field is zero in the setups we study here, we will refer only to the fluctuating part of the field $\vec b$.}
\begin{align}
  \partial_t \vec{v}  = q\Omega_0 x\partial_y\vec{v} - (\vec{v} \bm{\cdot} \vec{\nabla})\vec{v} -\nabla P + \vec{b}\bm{\cdot}\nabla{\vec{b}} + \nu \dif{\vec{v}}- \nonumber\\
  -2\vec{\Omega}_0 \bm{\times} \vec{v} + q\Omega_0 v_x \ey  \label{eq:mom},\\
  \partial_t \vec{b} =q\Omega_0 x\partial_y\vec{b}+ \vec{\nabla} \bm{\times} (\vec{v} \bm{\times} \vec{b}) + \eta \dif{\vec{b}} - q\Omega_0 b_x \ey \label{eq:induct},
\end{align}
In this system, the magnetic field is measured in the Alfv\'{e}nic units,~$v_A = b/\sqrt{4\pi\rho_0}$, the density $\rho_0$ is constant, and we use the dimensionless variables: the time is normalized to $t_0 = \Omega_0^{-1}$, the spatial variables to $L^{*}$, and the velocity to $\Omega_0 L^{*}$, where $L^{*}\equiv \min(L_x,L_y,L_z)$.
The dimensionless pressure $P$ is not an independent field, but rather is defined as to ensure the incompressibility of the flow.
In our numerical simulations we chose $\mr{Pm}= \nu/\eta = 1$.

We will solve Equations \eqref{eq:mom}-\eqref{eq:induct} in a rectangular box $(L_x, L_y, L_z)$, assuming periodic boundary conditions in the $z$- and $y$-directions, and shear-periodic boundary conditions in the $x$-direction~\cite[][]{umurhan2004}.
It is useful to discuss the conservation laws of Equations \eqref{eq:mom}-\eqref{eq:induct}.
In the non-rotating system, the MHD equations conserve the quadratic integrals of energy, cross helicity, and magnetic helicity if the external energy supply and energy dissipation are absent \cite[e.g.,][]{biskamp2003,tobias2011}.
In the MRI case, energy is injected into the system by the instability.
From Eqs.~\ref{eq:mom}~and~\ref{eq:induct} one then derives for the energy~\cite[][]{longaretti2010}:
\begin{equation}
  \frac{d}{dt}\left< \frac{v^2}{2} +\frac{b^2}{2} \right>={\tilde \alpha}-\nu \langle \vec{\omega}^2 \rangle -\eta \langle \vec{j}^2 \rangle ,
  \label{energy_balance}
\end{equation}
where ${\tilde \alpha}\nbs=\nbs q\Omega_0 \langle v_xv_y - b_xb_y \rangle$ is the energy injection rate, and the angular brackets denote an average over the box.
The conservation law of cross helicity has to be modified in the shearing box as:
\begin{equation}
  \frac{d}{dt}\langle \vec{v}\cdot\vec{b}+(2-q)\vec{\Omega}_0\cdot \vec{A} \rangle=
  -(\nu +\eta)\langle \vec{\omega}\cdot \vec{j} \rangle,
  \label{cross_helicity}
\end{equation}
and the conservation law for magnetic helicity remains unchanged:
\begin{equation}
  \frac{d}{dt}\langle \vec{b}\cdot\vec{A} \rangle=-2\eta \langle \vec{j}^2\rangle.
\end{equation}
In these equations we introduce the vorticity  $\vec{\omega}=\nabla {\times} {\vec v}$, the current density $\vec{j}=\nabla {\times}{\vec b}$, and the vector potential $\vec A$, where $\vec{b}=\nabla\times\vec{A}$.
In the presence of the magnetorotational instability and in the absence of dissipation, the energy grows while the cross helicity and magnetic helicity do not.

The system \eqref{eq:mom}-\eqref{eq:induct} is solved using the pseudo-spectral, incompressible version of the code \textsc{Snoopy}~\citep{lesur2007}.
\textsc{Snoopy} includes physical dissipation, while it uses a Fourier transform (\textsc{FFTW 3} library) and a low-storage third-order Runge-Kutta (RK3) scheme for time evolution of all the fields.
To employ the shear-periodic boundary conditions, \textsc{Snoopy} remaps the Fourier components of the fields periodically every $\Delta t_{\mr{remap}}=|L_y/(q \Omega_0 L_x)|$~(see the mathematical details of the procedure in ~\cite[][]{umurhan2004}).

In our work, we consider three simulation boxes with the following dimensions $(L_x, L_y, L_z) \in \{(1,4,4), (1,4,16), (2,4,1)\}$.
The numerical resolution in the first two cases is 128 points per unit length of $L_x$ and $L_z$, and 64 points per unit length of $L_y$.
In the third case, there are 512 points per unit length of $L_x$ and $L_z$, and 256 points per unit length of $L_y$.
Thus, the first and last cases have the same number of points in the vertical direction and are comparable after rescaling the unit of length.
The first two cases have a viscosity of $\nu = 1/5000$.
The third case has a viscosity of $\nu = 1/45000$.
A summary of the cases studied is given in Table~\ref{tab:cases}.
Note that the energy and transport parameter reported are volume densities of these quantities.
The transport parameter is defined in the standard way: $\alpha = \langle v_xv_y - b_xb_y \rangle/(q\Omega_0 L_z)^2$.
The values shown in the table are averages over some time intervals in the steady states (cases I, II, and IIIb), and over a short time interval in the decaying case IIIa, where the energy is approximately constant.
Cases I and II were initialized with $\vec{B}_0 = (0,0,B_0\cos(2\pi x/L_x))$ and a large-scale, white-noise velocity configuration, and then evolved until a steady state is reached.
Case III is the continuation of the zero-net-flux simulations discussed in our paper \cite[][case IV]{walker2016}.
\begin{table}
  \caption{Setups examined in this study.
    The final case (III) is decaying until $t\approx 600$, and the numbers given for $E$ and $\alpha$ are the values averaged over short time intervals around the indicated times.
  }
  \label{tab:cases}
  \centering
  \begin{tabular}{c| c| c| c| c| c| c| c}
    \hline
    \hline
    Case & $L_x \times L_y \times L_z$ & $\nu^{-1}$ & ${E}$ & ${\alpha}\, (\times 10^{-2})$ 
    \\
    \hline
    I   & $1 \times 4 \times 4$ & 5000 & 0.015 & 0.015 
    \\
    II & $1 \times 4 \times 16$ & 5000 & 0.15 & 0.0081 
    \\
    IIIa (t=222) & $2 \times 4 \times 1$ & 45000 & 0.017 & 0.28 
    \\ 
    IIIb ($t>800$) & $2 \times 4 \times 1$ & 45000 & 0.0015 & 0.0037 
    \\
    \hline
  \end{tabular}
\end{table}

\section{Results}
\begin{figure}
  \input{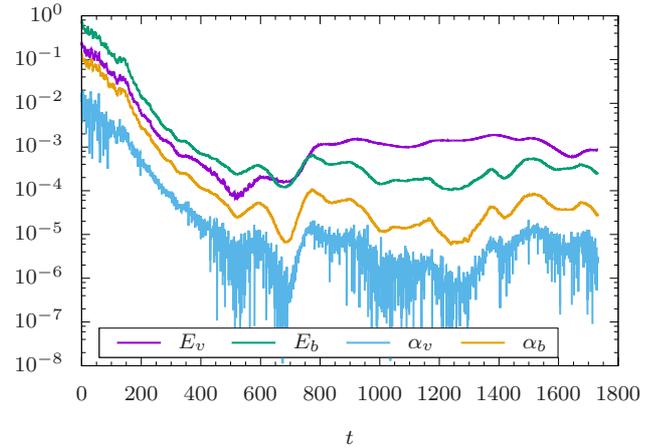}
  \caption{Time history for case III, showing the kinetic energy $E_v$, magnetic energy $E_b$, Reynolds stress $\alpha_v \equiv \langle v_xv_y \rangle/(q\Omega_0 L_z)^2$, and Maxwell stress $\alpha_b \equiv \langle -b_xb_y \rangle/(q\Omega_0 L_z)^2$.}
  \label{fig:time_history}
\end{figure}

Consistent with previous simulations~\citep{nauman2016}, we find that our boxes with the aspect ratio $L_z/L_x\ge4$  (cases I and II) are turbulent.
They rapidly reach a steady state and maintain it for the extent of the simulation, while $E$ and $\alpha$ fluctuate around a mean value.
Case III, which is a continuation of \cite[][case IV]{walker2016}, exhibits very slowly decaying turbulence until $t\approx 600$, and appears to reach a steady state at later times~(see Fig.~\ref{fig:time_history}).

\begin{figure}
  \input{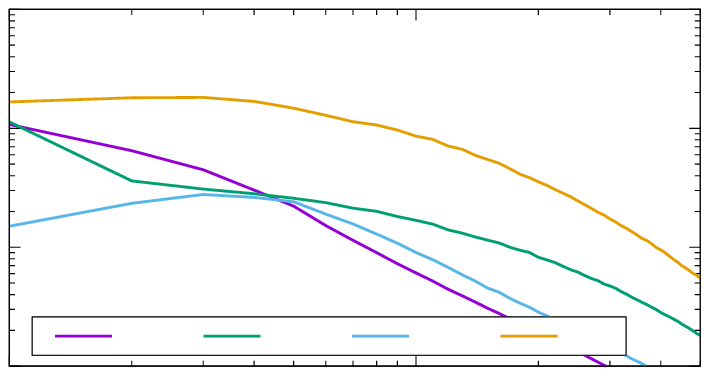}
  \input{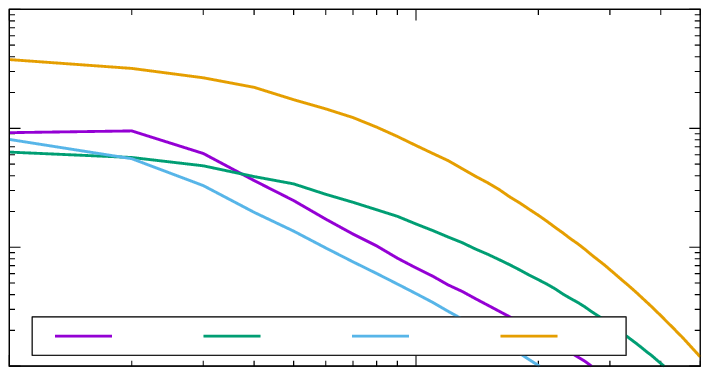}
  \caption{Energy spectra for case IIIa in the $x$-direction (upper panel) and $z$-direction (lower panel).}
  \label{fig:spectra_x2y4z1}
\end{figure}

In an attempt to explain these results, we find it useful to study the behavior of the azimuthal magnetic field $b_y$.
As was noted in our previous work \cite[][]{walker2016}, this field is concentrated at relatively large scales, and plays the role of the guide field for the remaining fluctuations.
We found, however, an important difference in the distribution of this field in tall and long boxes.
The net flux of the $b_y$ field is zero in the dynamo case.
Given some initial perturbation of the magnetic field, the instability sets in, which leads to stretching and folding of the magnetic field lines in the $x-y$ direction.
The shearing flow then increases the strength of the $b_y$ field.
An important observation, however, is that the $x$-scale of the resulting folds of the $b_y$ field is always on the $x$-scale of the fluctuating velocity field.
In the box extended in the $x$-direction (case III), this means that both the $b_y$ and the $v_x$ fields are still concentrated at comparable scales.
This is seen in the spectra shown in Fig.~\ref{fig:spectra_x2y4z1}.

\begin{figure}
  \input{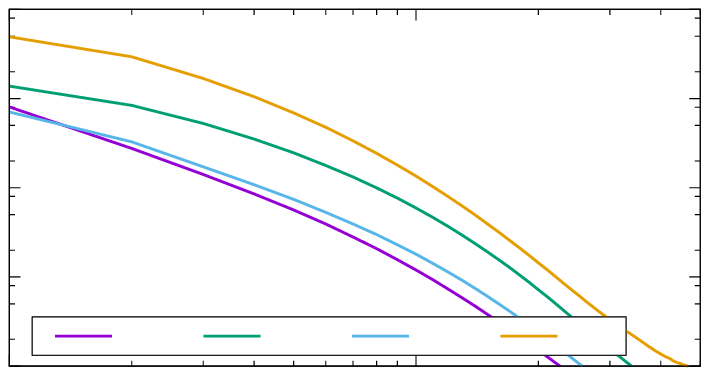}
  \input{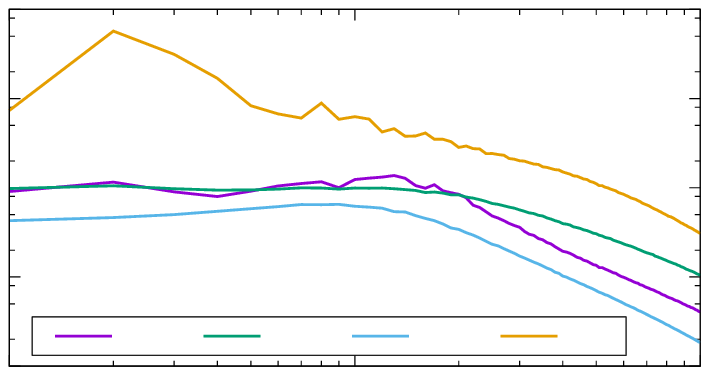}
  \caption{Energy spectra for case II in the $x$-direction (upper panel) and $z$-direction (lower panel).}
  \label{fig:spectra_x1y4z16}
\end{figure}

\begin{figure}
  \input{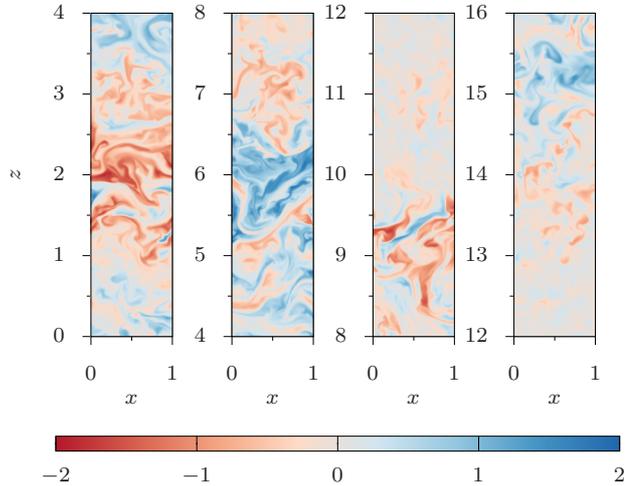}
  \caption{Snapshot of $b_y$ from case II at a late time when a steady state has been reached.
    Note that domain extends from 0 to 16 in the $z$-direction, but it has been partitioned here for ease of comparison.}
  \label{fig:snap_x1y4z16}
\end{figure}

\begin{figure}
  \input{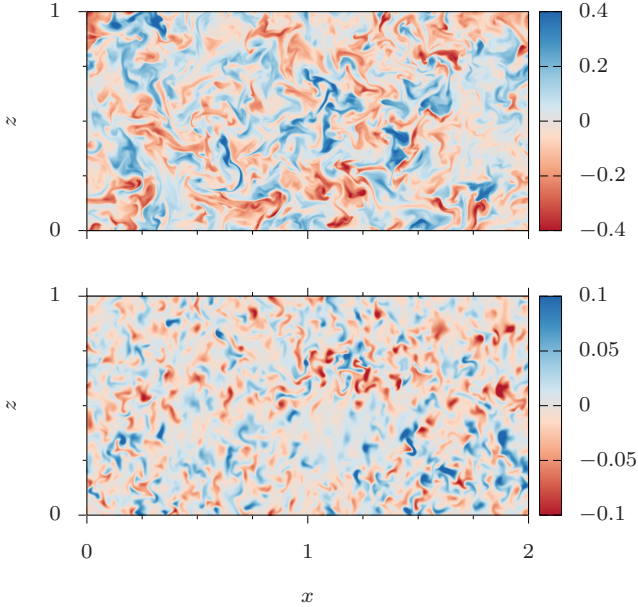}
  \caption{Snapshots of $b_y$ from case IIIa ($t=222$, top panel) and case IIIb ($t=907$, bottom panel).
    Note the decreasing scale of structures between IIIa and IIIb.}
  \label{fig:snap_x2y4z1}
\end{figure}

\begin{figure}
  \input{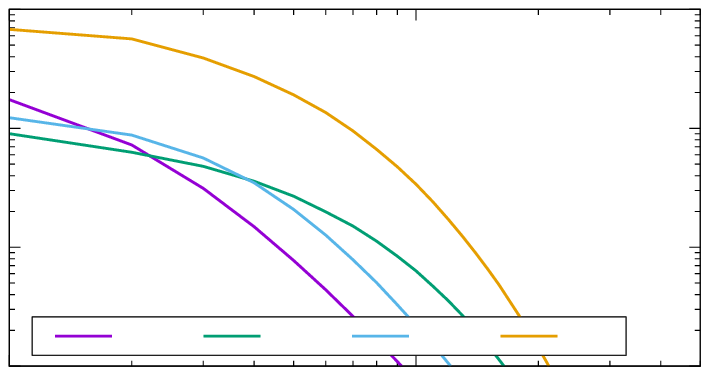}
  \input{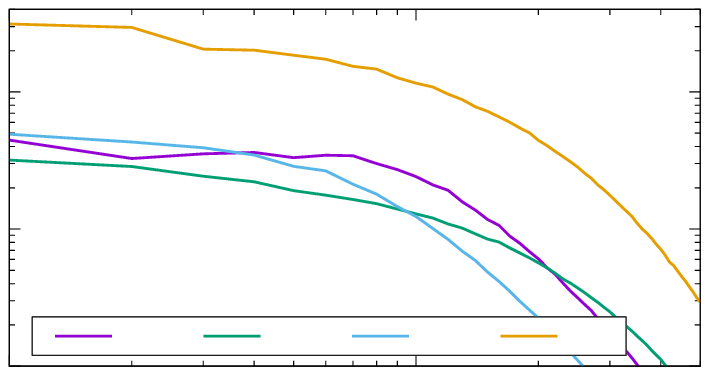}
  \caption{Energy spectra for case I in the $x$-direction (upper panel) and $z$-direction (lower panel).}
  \label{fig:spectra_x1y4z4}
\end{figure}

The situation is qualitatively different in a box extended in the vertical direction (case II).
It turns out that in this case the flux of $b_y$ can spread in the vertical direction over the scales much larger than the scales of the velocity field.
This is seen in Fig.~\ref{fig:spectra_x1y4z16}, where the $v_z$ component of the velocity field is concentrated at the scale comparable to the small horizontal box size $L_x$, while the $b_y$ magnetic field is concentrated at the larger scale $L_z$.
The structure of the $b_y$ field is shown in Fig.~\ref{fig:snap_x1y4z16}, which also reveals separation of the regions of positive and negative azimuthal magnetic flux in the vertical direction.
This is in contrast to Fig.~\ref{fig:snap_x2y4z1} where $b_y$ is more homogeneous, due to large-scale turbulent mixing.
Fig.~\ref{fig:spectra_x1y4z4} shows case I, an intermediate between cases II and~III.

\begin{figure}
  \input{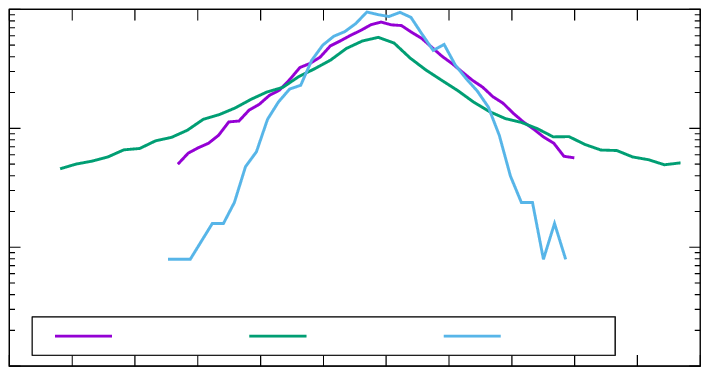}
  \input{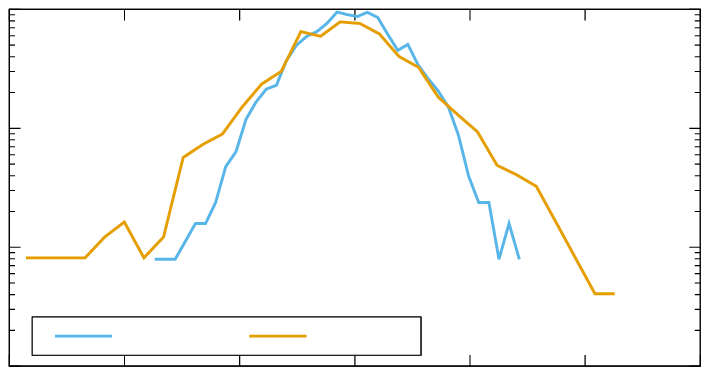}
  \caption{PDFs of flux $\Phi_y$ computed for each of the cases studied.
    Each dataset has been rescaled by the full width at half maximum $\Phi_{1/2}$.}
  \label{fig:histograms}
\end{figure}

In order to quantify this flux intermittency and to compare it with the other cases, we subdivided $y=\rm{const}$ cross-sections of the simulation boxes into small $1/16\times 1/16$ squares and calculated azimuthal magnetic fluxes for each of these squares.
We then plotted histograms of the obtained fluxes $\Phi_y$.
The results are shown in Fig.~\ref{fig:histograms}.
The flux intermittency is smallest in case IIIa, larger in case I, and the largest in the tallest box of case II.
The increased flux intermittency implies more favorable conditions for the dynamo action.

We are now in a position to answer why our case III did not exhibit dynamo action until the turbulence decayed to very small amplitudes~(see Fig.~\ref{fig:time_history}).
The matter is that when the energy of fluctuations declines, so does the scale of the turbulence~(see Fig.~\ref{fig:snap_x2y4z1}).
In our work \cite[][]{walker2016} this is explained by the necessity to maintain the balance between the linear and nonlinear shearing rates of the turbulence.
When the $x$-scale of the turbulence was relatively large compared to the box size $L_z$, the dynamo did not operate.
When the $x$-scale of the turbulence decreased as to become smaller than the $L_z$ size of the box, in analogy with cases II and I the dynamo action became possible.
Consistent with our discussion above, the intermittency of the magnetic flux also increased in this regime~(see Fig.~\ref{fig:histograms}).

\begin{figure}
  \input{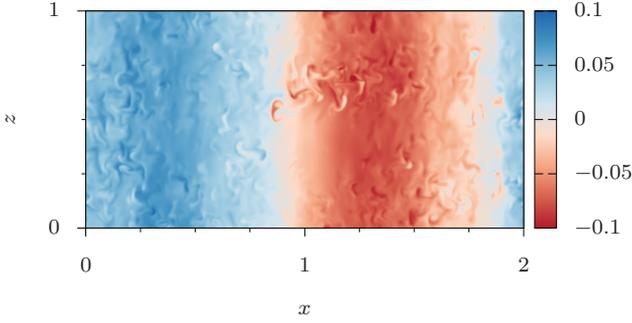}
  \caption{Snapshot of $v_y$ from case IIIb.}
  \label{fig:snap_x2y4z1_late}
\end{figure}

\begin{figure}
  \input{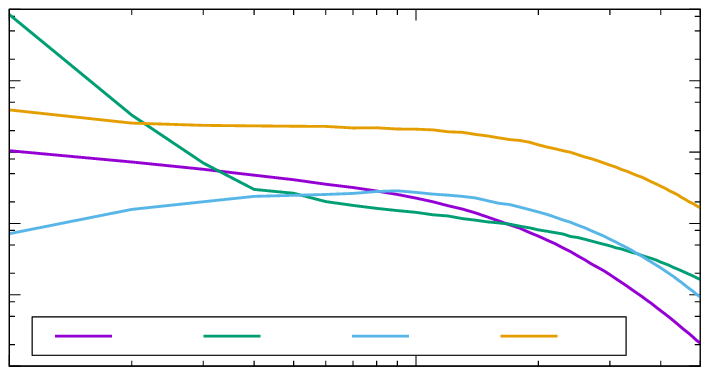}
  \input{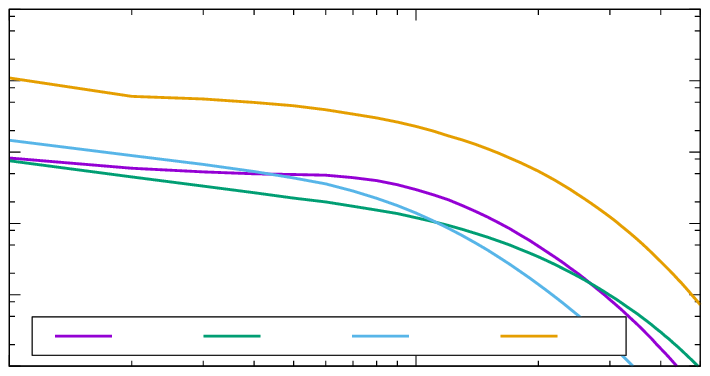}
  \caption{Energy spectra for case IIIb in the $x$-direction (upper panel) and $z$-direction (lower panel).}
  \label{fig:spectra_x2y4z1_late}
\end{figure}

Although the dynamo action seems to operate in case IIIb, we should exercise caution in claiming that a true steady state is observed.
As pointed out in~\citep{rempel2010} it may be possible in such a case that the system is described by a supertransient state and may ultimately decay if integrated long enough~\footnote{Although, with $\mr{Rm}=45000$, the expected lifetime is $\mathcal{O}(10^{32}$).}.
Also, unlike the other dynamo cases in this study, IIIb has developed, in addition to the small-scale fluctuations, a large-scale $v_y(x)$ zonal flow resulting from geostrophic balance~\citep{johansen2009}~(see Fig.~\ref{fig:snap_x2y4z1_late}). 
Comparing the spectra of case III from early times in Fig.~\ref{fig:spectra_x2y4z1} with those from late times in Fig.~\ref{fig:spectra_x2y4z1_late}, we see that this zonal flow is long-lived, or at least steadily reinforced. 
The possibility of such a flow may be related to the long radial extent of the box. 
This flow however is practically uncorrelated with the small-scale fluctuations; for instance, it does not contribute to the transport coefficient $\alpha_v$ in Fig.~(\ref{fig:time_history}).

\section{Conclusions}

We have found in incompressible MHD simulations of the shearing box that the properties of the zero-net-flux dynamo action depend intimately on the relation between the scale of the turbulence and the vertical size of the box.
For a given set of physical parameters, transport is sustained and a dynamo action rapidly sets in in systems with high $L_z/L_x$ aspect ratio, while in the opposite case of small aspect ratio, the dynamo action is not observed until the scale of the turbulence significantly decreases.
This is the despite the size of the box being much greater than the dissipative scale of turbulence in both cases.
Based on our results we suggest the following explanation for this phenomenon.
We propose that an inherent property of the shearing-box turbulence is its tendency to build up local regions of nonzero $b_y$ flux,  separated in the vertical direction $z$.
This may be related to the fact that the configurations ${\vec b}=(0, b_y(z),0)$, ${\vec v}=(v_x(z,t), v_y(z,t), 0)$ are exact solutions of the ideal equations \cite[e.g.,][]{goodman1994}.
Those sub-regions with nonzero $b_y$ fluxes are MRI-unstable.
The resulting turbulent eddies, on the other hand, are trying to mix and homogenize the $b_y$ field.
The size of the turbulent eddies is always  comparable to the size of the folds of the $b_y$ field in the $x$-direction.
Indeed, such eddies are caused by the MRI instability of the $b_y$ field.
In ``long'' boxes, the eddies caused by an initially unstable large-scale perturbation have a typical length-scale larger than the vertical extent of the box, and they are able to effectively mix the field in that direction, thus reducing the $b_y$ fluxes and effectively reducing the rate of the instability as compared to the rate of turbulent energy dissipation.
The resulting turbulence will therefore decay, until the scale of the turbulence becomes smaller than the vertical size of the box, at which point the vertical flux separation enhances the efficiency of the instability.
In ``tall'' boxes, on the other hand, the eddies are always smaller than the vertical extension of the box.
This allows oppositely-directed $b_y$ fluxes to build up in vertically separated regions, allowing the instability to win. 

This poses a broader question to what extent the shearing-box results describe the natural MRI turbulence.
The shearing-box model suggests that the scale of the MRI-dynamo-driven turbulence should be smaller than the thickness of the disc and that the resulting transport coefficients are very small.
We, however, notice that the vertical limitations in real accretion disks are never as severe  as the periodic boundary conditions of the shearing-box simulations.
Allowing the flux of the $b_y$ magnetic field to partially escape the box in the vertical direction, for instance, will facilitate the dynamo action rather than impede it.
This indicates that the shearing-box model with periodic boundary conditions may not allow one, at least at the present level of understanding, to realistically simulate the scale and the transport properties of the MRI-dynamo-driven turbulence at low magnetic Prandtl numbers.
What seems to be a robust feature of the MRI-dynamo-driven turbulence, however, is its tendency to separate the regions of positive and negative $b_y$ fluxes, leading to highly intermittent azimuthal magnetic flux patches.
In a stratified disk, these magnetic flux tubes can then buoy above the surface, forming an intermittently magnetized corona~\cite[e.g.][]{galeev1979}.
Non-periodic large-scale conditions (e.g., stratification, flux escape, etc.) need to be incorporated in the shearing-box model in order to provide a more realistic description. 

\section*{Acknowledgments}
We thank Fausto Cattaneo and Geoffroy Lesur for useful comments.
JW is supported by the Wisconsin Alumni Research Foundation at the University of Wisconsin-Madison.
SB is partly supported by the National Science Foundation under the grant NSF AGS-1261659 and by the Vilas Associates Award from the University of Wisconsin-Madison. Simulations were performed at the Texas Advanced Computing Center (TACC) at the University of Texas at Austin under the NSF-Teragrid Project TG-PHY110016.

\bibliographystyle{mnras}
\bibliography{plasma}

\end{document}